\documentclass{aastex63}
\usepackage{epsfig}                     
\usepackage{graphicx}                    
\usepackage{courier}     
\usepackage{amssymb}                    
\usepackage{color}                       
\graphicspath{{./}{figures/}}

\begin{document}

\title{On the short term modulation of cosmic rays by high-speed streams  at the Pierre Auger surface array detectors}

\correspondingauthor{C.E. Navia}
\email{navia@if.uff.br}

\author{M.N. de Oliveira}
\affiliation{Departamento de Geof\'isica, Observat\'orio Nacional, 20921-400, Rio de Janeiro, Brazil}

\author{C.R.A. Augusto}
\affiliation{Instituto de F\'{i}sica, Universidade Federal Fluminense, 24210-346, Niter\'{o}i, RJ, Brazil }

\author{C.E. Navia}
\affiliation{Instituto de F\'{i}sica, Universidade Federal Fluminense, 24210-346, Niter\'{o}i, RJ, Brazil }


\author{A.A. Nepomuceno}
\affiliation{Departamento de Ci\^encias da Natureza, Universidade Federal Fluminense, 28890-000, Rio das Ostras, RJ, Brazil}

\begin{abstract}
We present an analysis of the short-term modulation (one rotation of Bartels-27 days) of the galactic cosmic rays (GCR)  by the solar wind, based on the cosmic ray rates observed by the Pierre Auger Observatory (PAO)  on their surface detectors in scaler mode. The incidence of GCR with energies below $\sim$ 50 TeV, at the top of the atmosphere, produces more than 90\% of the secondary particles registered at ground level, i.e., they are subject to solar modulation.
The modulation is consistent with at least two components: The first is the modulation of the amplitude of the cosmic rays diurnal variation, anti-correlated with the solar-wind speed. The second one occurs during the high-speed stream (HSS),  the baseline of the cosmic rays diurnal variation train falls,
following the time profile of the solar-wind speed inversely.
Based on the radial gradient of the cosmic ray diffusion theory and under some other premises, such as the latitude dependence on diurnal variation and the inclusion of drift processes in the propagation of GCR, a semi-empirical description of the modulation is possible to do, and it hereafter is called as Toy-model.
  Although the Toy-model does not include fluctuations due to propagation in the atmosphere, it provides satisfactory results when compared with
the PAO scaler mode data. We present details of these observations as well as the Toy-model validation.
\end{abstract}


\section{Introduction} 
\label{sec1}

The discovery of ionizing radiation coming from space by \cite{hess12} is the beginning of research on cosmic rays. The origin, composition, energy spectrum, and propagation allows us to obtain valuable information on the Earth's environment, the heliosphere, the galaxy, and in the ultra high energy, it is possible to probe, extra-galactic information. The cosmic rays research plays a vital role in the field of space physics and remains a very active area. 

The cosmic ray research led to the birth of particle physics.
The first evidence of antimatter, predicted by Dirac quantum relativistic theory, was in 1932. \cite{ande33} discovered antimatter in the cosmic radiation, in the form of the anti-electron, later called the positron. A new particle, now called a muon, also was found by \cite{ande36} within the cosmic radiation. Initially, it was believed that this particle would be the mediator meson of the nuclear forces. However, the discovery of meson Pi predicted by the Yukawa nuclear theory was only in 1947 by \cite{latt47}. This discovery opens the door to the study of nuclear collisions using cosmic rays induced events. Indeed, starting in 1960, the nuclear emulsion chambers on a large scale are used to study the nuclear collisions at high mountain (Chacaltaya-Bolivia)\citep{latt80}.  Results obtained by Brazil-Japan collaboration were researchers later by the SppS at CERN 
\citep{rush81}.

From photographs of the cloud chamber of events induced by cosmic rays, \cite{roch47} obtained the first evidence of strange particles. It is the discovery of meson K.
 Also, \cite{niu71} had the first evidence of long-flying particles in events induced by cosmic rays, later identified as charming particles.
 Thus, of the six particle flavors of the standard model, four of them
were discovered in cosmic radiation.

\cite{auge39} using two detectors with a distance of many meters between them noticed that they were triggered simultaneously,
  indicating the arrival of particles at the same time. It is the discovery of the particle shower. The incidence of a high-energy cosmic ray in the atmosphere produces a shower of secondary particles.

On the other hand, the first information about the diurnal variation in the flow of cosmic rays was made by \cite{forb37}.
However, large-scale observation of daily variation began only in the 1950s \citep{elli52}. In these years, Simpson builds the first Neutron Monitor (NM), subsequently setting up a network of these NMs on the three American continents. The network has been extended to other continents, since the International Geophysical Years (IGY). NMs are one of the best cosmic ray detectors, in the energy range from 0.1 to 1000 GeV at great atmospheric depths (ground level).

The century 21, is the beginning of the multi-messenger astronomy 
and will allow us to understand more accurately the physics of the most violent events in the universe, such as supernovae and fusion of neutron stars and others.
Multi-messengers, such as the ultra-high energy (UHE) cosmic rays, $\gamma$-ray bursts, neutrinos, and gravitational waves,  will allow a better understanding of these events, and up to maybe find a new physics.

In 1995 an international group of researchers under the leadership of Cronin \citep{auge95} began design studies for a new cosmic ray observatory, the Pierre Auger Project. In 2004 the Pierre Auger  Observatory (PAO),  using a giant hybrid detector, start the observations of the ultra high-energy cosmic rays \citep{auge04}. So far, several results were obtained, such as the chemical composition \citep{Aab17}, the confirmation of the GZK cutoff in the ultra high energy region of the cosmic ray spectrum \citep{sett12}. The existence of a dipolar anisotropy on the very-high-energy cosmic ray sources \citep{aab18}, giving valuable information on the origin and evolution of the universe.

The main objective of this article is the study of the modulation by the solar-wind of the cosmic rays diurnal variation amplitude, and the high-speed streams, modulating the baseline of the diurnal variation train. On the base of PAO scaler mode data, and the radial gradient of the cosmic rays diffusion theory.

 With this aim, the paper organization follows as:
Section 2 is a brief description of the PAO scaler mode data.
In section 3, the application of a Fast Fourier Transformation (FFT) in the hourly rates of the PAO data, covering almost 13 years, shows at least six types of anisotropies.  Of them, the diurnal anisotropy is the one that presents a higher confidence level. 
Section 4 is devoted to a description of high speed-streams, mainly from coronal holes, but also includes some information from HSSs from active solar regions, linked with solar flares.

Section 5 is devoted to a formulation of a semi-empirical model, called Toy-model based on the radial gradient of the diffusion theory. Section 6 includes some qualitative comparisons between the 
Toy-model predictions and PAO data, as well as quantitative analysis, i.e.,  the Toy-model validation. The description of the solar wind-speed data comes from the Advanced Composition Explorer (ACE) spacecraft at Lagrange Point L1. 

Section 7 includes the analysis of the effects on the cosmic rays of an HSS from coronal mass ejection. A brief description of the long term modulation of the GCR by the solar-wind speed is present in section 8, and finally, in section 9, we present our conclusions.


\section{Escaler mode data from PAO}

The PAO  localization is in Malargue,
Argentina (35$^{\circ}$S; 69$^{\circ}$W, 1400 m a.s.l.). The goal
is the study of cosmic rays at the highest energy region above EeV (10$^{18}$ eV). The construction of a Hybrid Detector provides two independent methods to measure the energy, chemical composition, and arrival direction of UHE cosmic rays by cross-calibration.

The Surface Detector Array detects showers of secondary particles produced by collisions of very high energy cosmic rays with nuclei in the upper atmosphere.  The detector consists of 1660 water-Cherenkov detector stations, arranged in a 1500 m spaced triangular grid covering a total area of 3000 km$^2$. A trigger system allows us to obtain the arrival direction of the primary particle, and the shower size allows us to estimate the primary particle energy.

Also, a fluorescence light produced by collisions of very high energy cosmic rays with the nitrogen in the upper atmosphere can be detected in moonless nights by the fluorescence detector composed of 27 telescopes housed at four different locations at the edges of the Surface Detector array.

Shower fluorescence images allow us to reconstruct the cascade's curve and to obtain the mass composition and energy of the primary particle and to make a cross-check with the measurements of the surface array.

On the other hand,  the scaler mode, counts all the particles hitting the detectors of the surface array, without any trigger system \citep{auge11}.
The PAO data is public and available through the site \url{http://www.auger.org}. The scaler mode data, 15 minutes counting rates, averaged over all water Cherenkov detectors,  is available to download from March 1, 2005. 

As the geomagnetic rigidity cutoff at Malargue is 9.5 GV, incident primary cosmic rays in the upper atmosphere with energies between 10 GeV to 2 TeV produce 90\% of the counting rate in one unit (water Cherenkov detector) of the PAO surface array \citep{dass12}.
Cosmic rays in this energies range are subject to solar modulation. Indeed, the effect of transient solar events such as
the Forbush decrease  during geomagnetic storms was reported using the scaler mode data of PAO
 \citep{auge11}. As well as other correlations, with the sunspots   \citep{cana12}.

Also, even after atmospheric corrections, a diurnal variation with an average amplitude of 0.25\% remains in the PAO data. That is the main solar modulation in the PAO scaler mode data, as well as in several ground-level detectors, such as the neutron monitors (NMs).

\section{Galactic cosmic-ray anisotropies at 1 AU}
\label{galactic_ani}

\begin{figure*}[th]
\vspace*{-0.0cm}
\hspace*{-0.0cm}
\centering
\includegraphics[clip,width=0.9
\textwidth,height=0.4\textheight,angle=0.] {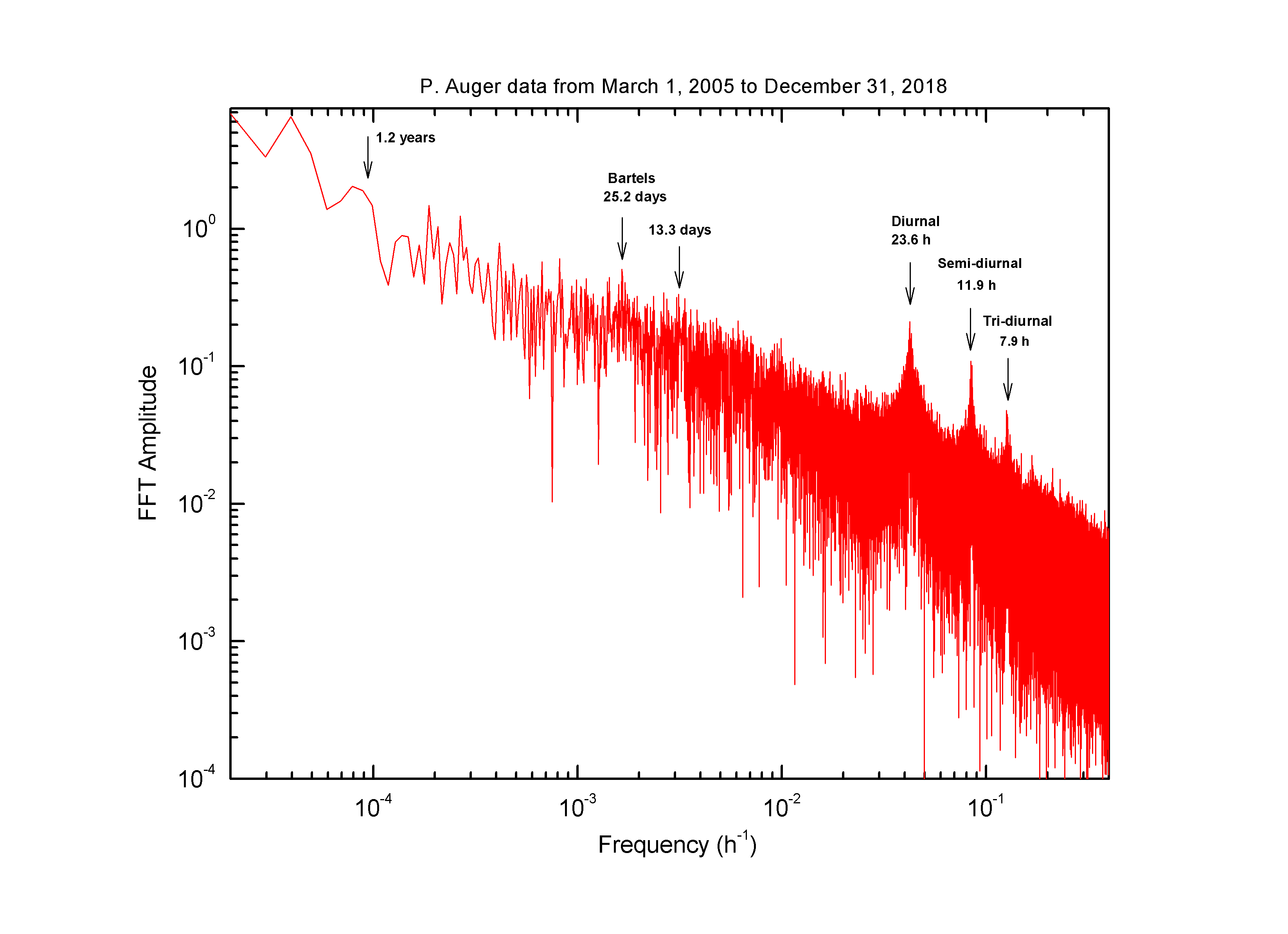}
\vspace*{-0.5cm}
\caption{Fast Fourier Transformation amplitude as a function of the frequency, obtained from the hourly particle rates on the PAO scaler mode data, from March 1, 2005, to  December 31, 2018.
}
\label{power}
\end{figure*}

The observed anisotropies in the galactic cosmic-rays, with energies below 50 TeV, mean that their propagation in the inner heliosphere is not isotropic. There are parallel and perpendicular gradients to the ecliptic plane. They are responsible for the solar diurnal anisotropy, as well as, responsible for the North-South anisotropy and observed at the Earth as a semi-diurnal anisotropy, among others. The hourly rates, from almost 13 years of the PAO scaler mode data, were used to obtain the Fast Fourier Transformation (FFT). The result is shown in   Fig~\ref{power} and represents the amplitude of the FFT as a function of frequency.

At the FFT-plot, we can observe that the main one is the diurnal solar anisotropy, with a confidence level of 69.2\% and periodicity of 23.6 h.
The second one is the semi-diurnal anisotropy, with a confidence level of 53.8\% and periodicity of 11.9 h, the third one,  is the tri-diurnal anisotropy with a periodicity of 7.9 days and confidence level of 45.4\%. The PAO data set, also permit to see the Bartels anisotropy (due to sun rotation), with a periodicity of 25.2 days and a confidence level of 34.2\% while the semi-Bartels anisotropy (periodicity $\sim 13.3$ days) is seen only marginally.

 However, the annual anisotropy is already glimpsed. The observation of the modulation with the variation of the 11-year sunspot cycle requires more long-term data.

\section{Coronal holes and high speed streams}

Coronal holes are colder regions with lower density plasma in the solar corona and appear as dark areas in the polar regions and close to the equator \citep{rich18}.
The coronal holes correspond to regions of open magnetic field lines, allowing the solar wind to escape more quickly to the interplanetary space. That is, the primary source of a high-speed stream is the coronal holes (CH), and there are at least two by Bartels rotation. Besides, solar flares and CMEs give rise to high-speed flows; in most cases, they are more intense, but of shorter duration and occur mainly during the maximum sunspot cycle.

During the high solar activity phase, the coronal orifices exist at all latitudes, but are less persistent, with a lifetime of less than two Bartels rotations. Already during the declining solar phase and close to the solar minimum, they are located in the polar regions, but
they extend at low latitudes and are more persistent with a lifetime of several Bartels rotations.

Coronal Holes and HSSs are closely related; the stream formation is on the coronal hole region. This high-speed solar wind follows the open magnetic field lines into the heliosphere.
Not all high-speed streams generated from coronal holes reach the earth. In most cases, only those streams at or near the ecliptic plane have geo-effectiveness. So, HSSs from low latitude coronal holes follow the
interplanetary magnetic field spiral pattern (Parker spiral).


\begin{figure}[th]
\vspace*{+0.0cm}
\hspace*{-1.0cm}
\centering
\includegraphics[clip,width=0.5
\textwidth,height=0.4\textheight,angle=0.] {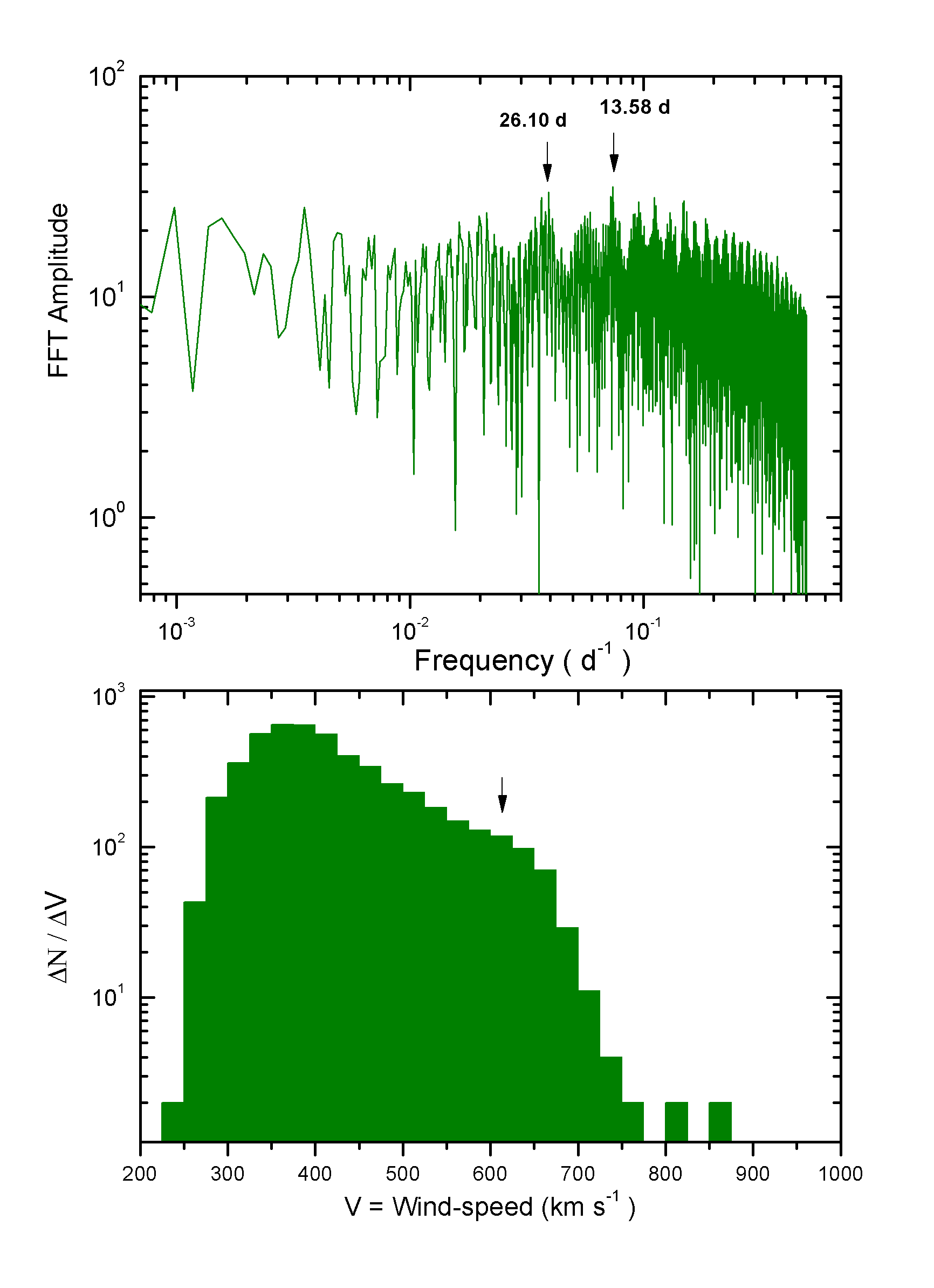}
\vspace*{-0.5cm}
\caption{Top panel: Fast Fourier Transformation  amplitude as a function of the frequency, obtained from ACE-SWEPAM solar wind speed averages rates at L1, from January 1, 2000, to December 31, 2013. Bottom panel: Solar wind speed distribution, the data correspond to the same period above shown. The vertical arrow indicates that for wind speed streams about or higher to 600 km/s the streams can trigger geomagnetic storms.
}
\label{FFT_wind}
\end{figure} 

\begin{figure}[th]
\vspace*{+0.0cm}
\hspace*{-1.0cm}
\centering
\includegraphics[clip,width=0.5
\textwidth,height=0.4\textheight,angle=0.] {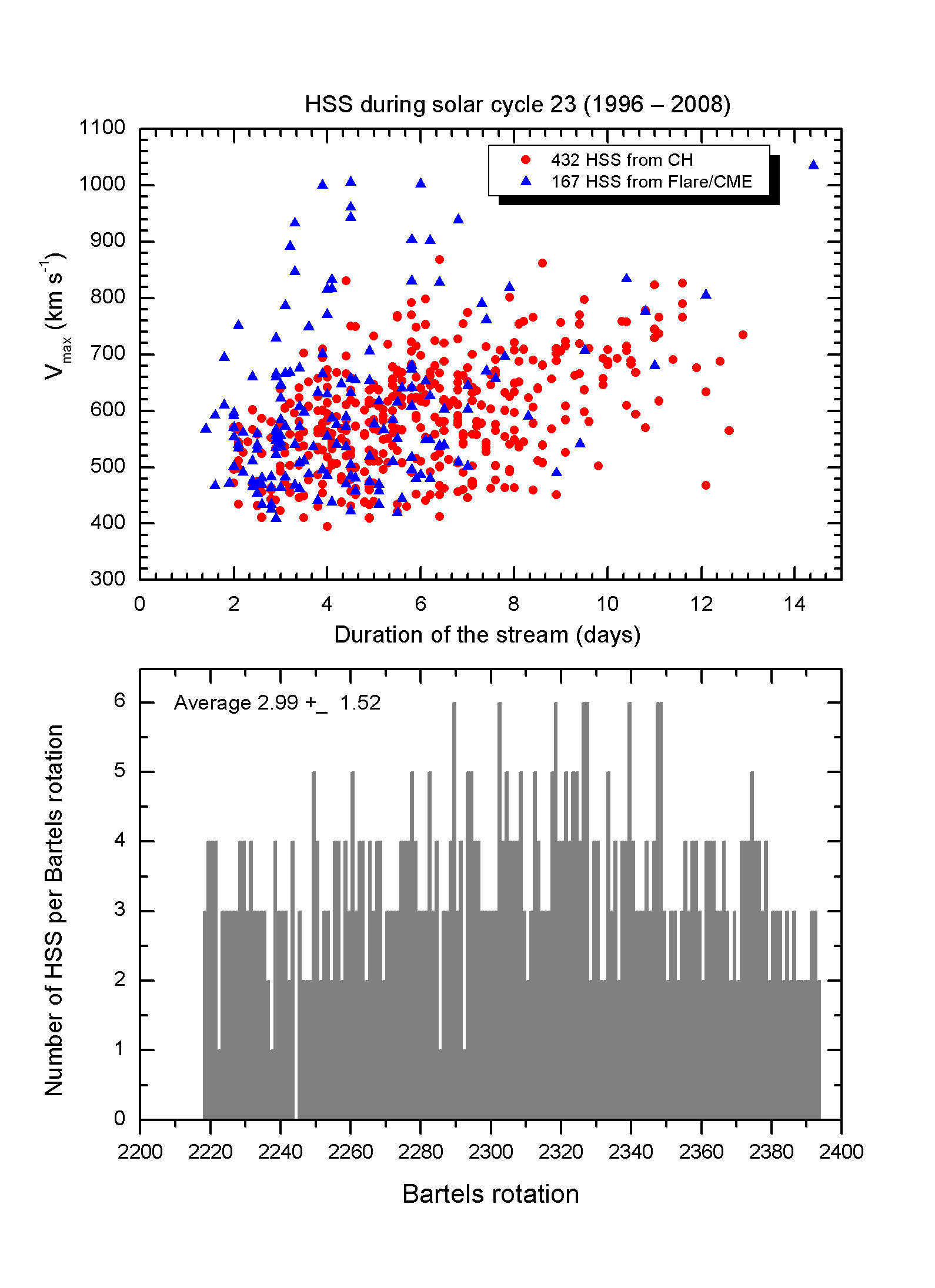}
\vspace*{-0.5cm}
\caption{Top panel: Correlation between the solar wind stream duration and the maximum speed of the solar wind. Bottom panel: Distribution of the number of solar wind streams, only of coronal holes, by Bartels rotation. In both panels, the data is from
SOHO CELIAS / MTOF Proton Monitor onboard the SOHO probe at L1, during the entire solar cycle 23.
}
\label{soho}
\end{figure} 

To observe if there is any periodicity in the solar wind flow that reaches the terrestrial environment, we performed a Fast Fourier Transform (FFT) in the daily solar wind averages obtained from the SWEPAM-ACE data at the Lagrange L1 point. The data cover a period of 13 years, from January 1, 2000, to December 31, 2013, and include a minimum solar cycle period, 2008 and is available in \url{http://www.srl.caltech.edu/ACE/ASC/}. The top panel in Fig.~\ref{FFT_wind} shows the result. 

Following  Fig.~\ref{FFT_wind}, we can see that there are two low significance signals with periodicities of $\sim$ 26.10 days and $\sim$ 13.58 days  (one and two HSSs per Bartels rotation, respectively). A sign with a periodicity of  $\sim$ 9 days is seen only marginally.
Already the bottom panel of Fig.~\ref{FFT_wind} shows the solar wind distribution. As expected, the mean velocity is around 
400 km s$^{-1}$. However, there is a kind of a fall in the distribution above 600
km s$^{-1}$. Streams with this speed or higher (marked by a vertical arrow), when reaching the magnetosphere, can trigger geomagnetic storms.

More information about HSSs, such as duration and multiplicity per Bartels rotation, can be obtained by analyzing the full data from the last
solar cycle 23, obtained by SOHO CELIAS/MTOF Proton Monitor onboard the SOHO probe on L1. The high-speed catalog (1996-2018) was compiled by
Maris, Maris and available in \url{http://www.spacescience.ro/00-old/new1/HSS_Catalogue.html}

During the solar cycle 23, the SOHO CELIAS/MTOF registered 432 HSSs from coronal holes and 163 HSSs from flare/CMEs. The top panel in Fig.~\ref{soho} shows a correlation between the maximum speed of the HSS and the duration.
We can see that the HSSs from coronal holes tend to have a significant duration than the HSS  from flare/CMEs. However, the HSSs from flares/CME tend to be more intense (higher $V_{max}$) than the HSSs from coronal holes.
Already the bottom panel in Fig.~\ref{soho} shows a  distribution of the number of HSSs per Bartels rotation from CH only. Following the figure, we can see that there are at least two HSSs per Bartels rotation with an average value of  2.99 $ \pm $ 1.52.   Also, the number of HSSs during the minimum solar cycle tends to be a little smaller than during the maximum.



\section{Solar-wind speed modulating the GCR}
\label{modulation}

The Fokker-Plank transport equation to the case of galactic cosmic ray propagation in the inner heliosphere was developed by \cite{park65}, in terms of parallel and perpendicular diffusion components to the mean magnetic field.
The Parker equation is quite complicated, and in most cases, there are only some analytical solutions restricted to special conditions and approximations. A most robust solution requires numerical methods. However, some limitations of the theory are due to the uncertainties of determining the transport coefficients for diffusion \citep{fisk80}. 
They would be not constants but would have dependence, with the energy and with the space position on the heliosphere.

\subsection{Solar diurnal anisotropy}
\label{diurnal}

Galactic cosmic-rays propagation (with rigidities below to $\sim$50 GV) in the ecliptic plane corotates with the interplanetary magnetic field (IMF).  When the flow reaches the Earth, produces the solar diurnal anisotropy, roughly perpendicular to the Earth-Sun line, direction $\sim$ 18 h LT, observed by a detector on the spinning Earth. 
 \begin{figure}[th]
\vspace*{-0.0cm}
\hspace*{-1.0cm}
\centering
\includegraphics[clip,width=0.35
\textwidth,height=0.4\textheight,angle=0.] {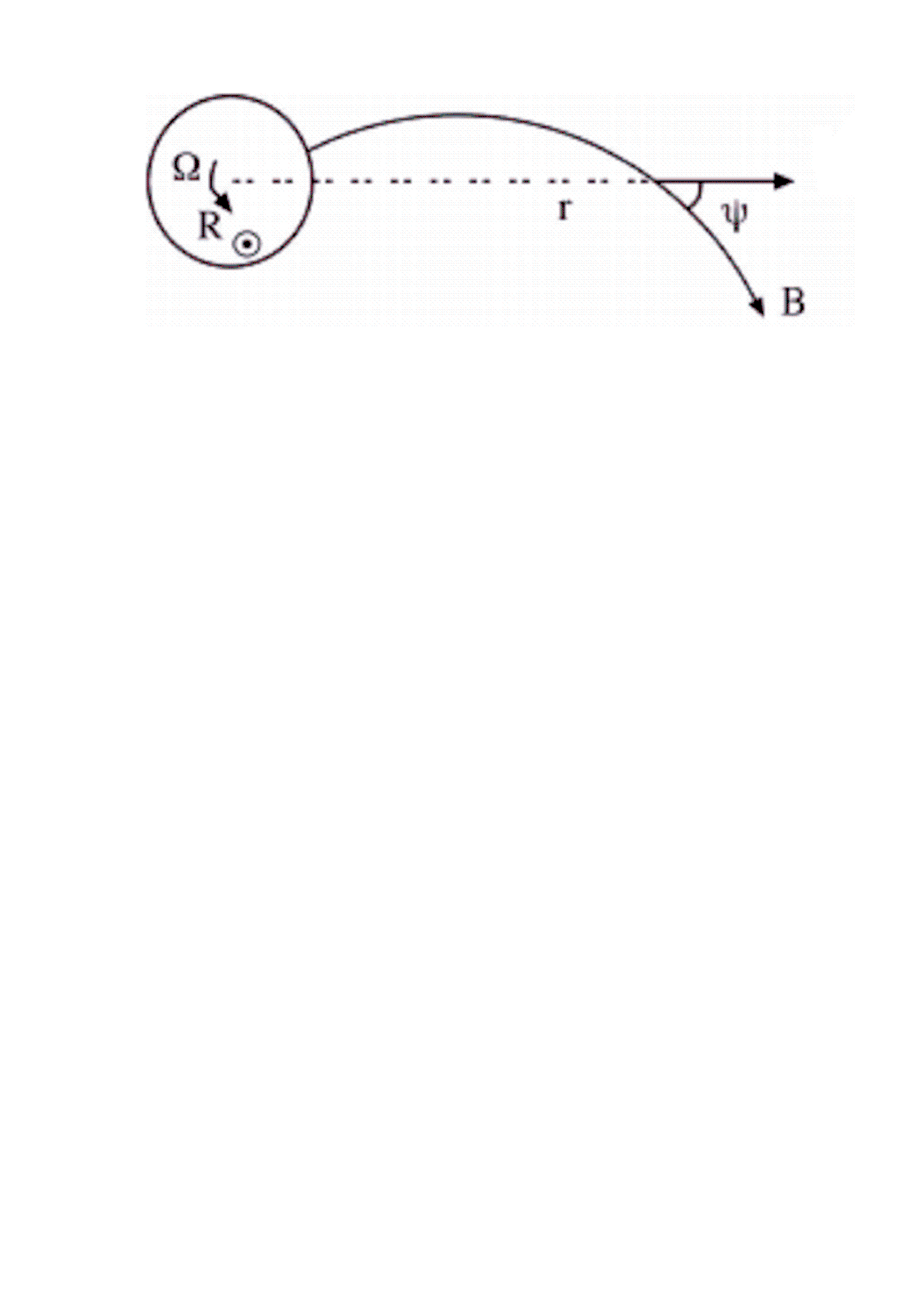}
\vspace*{-6.0cm}
\caption{Definition of the angle $\Psi$, as the angle between the mean magnetic field and the Sun-Earth radial direction, $r$. 
}
\label{phi_angle}
\end{figure} 

 Lets beginning indicating the relation for the solar magnetic field that executes a spiral pattern at large radial distances (r) \citep{fisk80}. It is a relation  among three variables, the angle $\Psi$ that is the angle between the magnetic field direction and the radial direction, the Sun-Earth line, as shown
 in Fig.~\ref{phi_angle}, the solar wind-speed (V) and the polar angle  ($\theta$) that define the sun latitude, the relation under the assumption of no drift effects yields
\begin{equation}
\cos^2 \Psi=\frac{V^2}{V^2+\Omega^2 r^2 \sin^2 \theta},
\label{fisk}
\end{equation}
where $\Omega$ is the angular speed of the sun ($\Omega=2.7 \times 10^{-6}$ rad s$^{-1}$) and
  $r=1.496\times 10^{8}$ km (mean Sun-Earth radial distance). If $\theta=\pi/2$ the particle motion is restrict only to the ecliptic plane, and Eq.~\ref{fisk} with this restriction become
\begin{equation}
\tan \Psi=\frac{\Omega \; r}{V}.
\label{tan_psi}
\end{equation}

\begin{figure}[th]
\vspace*{-0.0cm}
\hspace*{-1.0cm}
\centering
\includegraphics[clip,width=0.5
\textwidth,height=0.5\textheight,angle=0.] {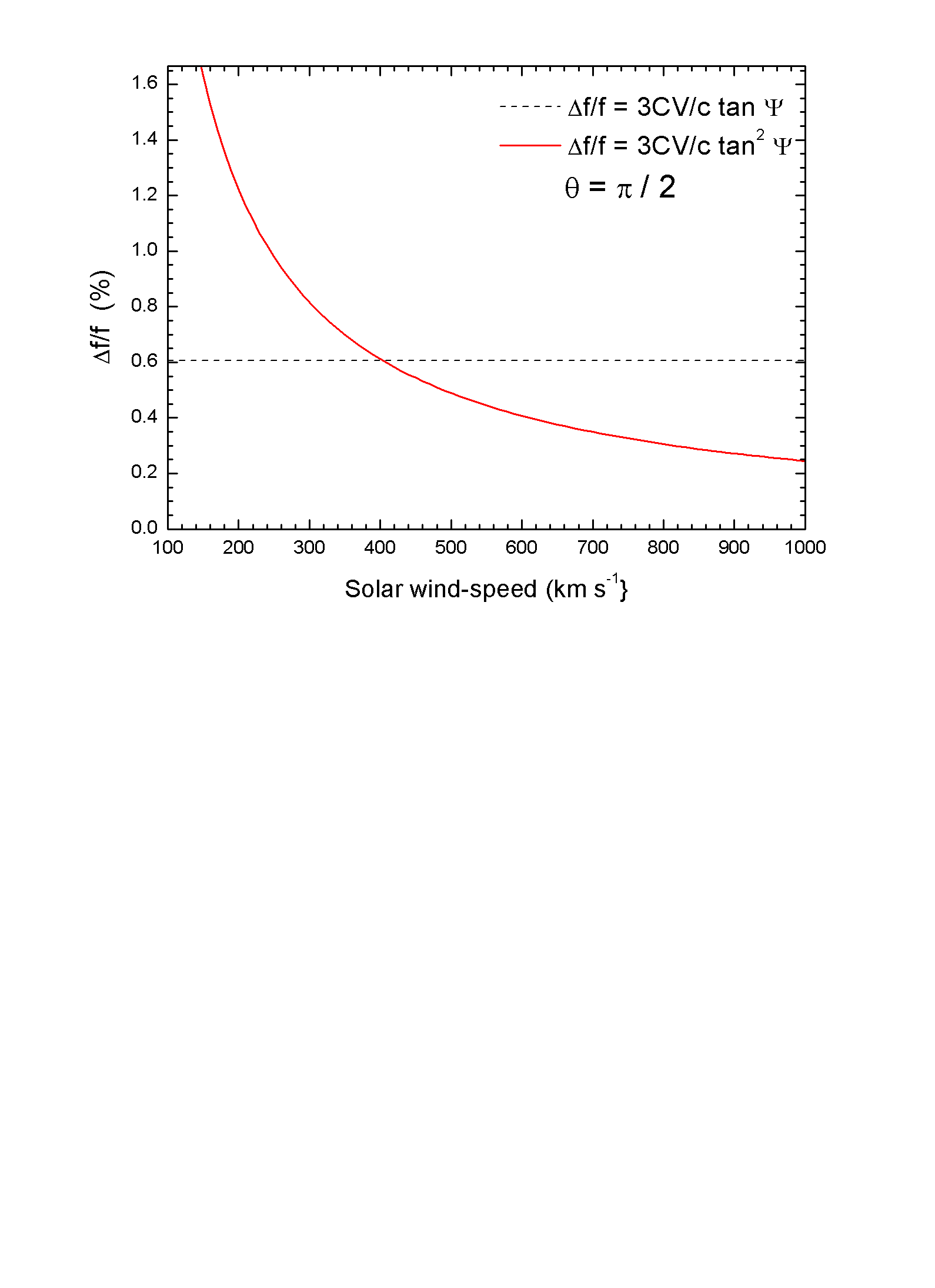}
\vspace*{-6.0cm}
\caption{Dependence of the amplitude of the solar diurnal variation with the solar wind speed. The dotted line represents the first approximation of the diffusion theory (Eq.~\ref{parker}). The solid curve represents the azimuthal component of the radial gradient of the diffusion theory
(Eq.~\ref{ani1}).
}
\label{corotating}
\end{figure} 

The diffusion theory can describe the propagation of cosmic rays in the inner heliosphere. It explains the diurnal solar anisotropy
\citep{park64,park67,axfo65,urch72,glee67} .
 Initially, the theory provided an approximation considering
a zero radial flow and in the limit $K_{\perp}/K_{\parallel}<<1$, where 
$K_{\perp}/K_{\parallel}$ is the ratio of perpendicular to parallel diffusion coefficients.
In this approximation, the  amplitude of the diurnal anisotropy yields
\begin{equation}
\frac{\Delta f}{f}=A_{\phi}=3C \frac{V}{c}\tan  \Psi= 3C \frac{\Omega r}{c},
\label{parker}
\end{equation}
here, C is the Compton-Getting factor (C$\sim$ 1.5) to particles with energies above GeV, and in this energy region, the GCR propagation has a speed close to c (light-speed).
Eq.~\ref{parker} can be called as a first approximation, because it predicts an amplitude to the diurnal anisotropy, independent of solar wind speed as $A_{\phi} =0.6\%$. The dashed black line in Fig.~\ref{corotating} represents this solution. However, the data show a modulation of the amplitude.


\subsection{Modulation of amplitude and train variation}

A second approximation for the diurnal variation amplitude is built-in the azimuthal solution of the radial gradient of diffusion theory
\citep{rike87} as
\begin{equation}
\lambda_{\parallel}G_r= -A_r +3C \frac{V}{c}+A_{\phi} \tan \Psi.
\label{riker}
\end{equation}
Where $A_r$ and $A_{\phi}$ are the radial and azimuthal components of the diurnal anisotropy, respectively. The radial component $A_r$ is about ten times smaller than
the azimuthal component $A_{\phi}$ \citep{ahlu15} 
and the last terms on the right of Eq.~\ref{riker} constitute the azimuthal component. Incorporating  Eq.~\ref{parker} into Eq.~\ref{riker},  we have a expression to the diurnal variation as
\begin{equation}
\frac{\Delta f}{f}=3C\frac{V}{c}\tan^2 \Psi,
\label{ani1}
\end{equation}
The solid (red) curve in Fig.~\ref{corotating} represents this solution. So, Eq.~\ref{ani1} predicts modulation of the amplitude of the diurnal variation in anti-correlation with the solar wind speed, as is observed on the data.

\begin{figure}[th]
\vspace*{-0.0cm}
\hspace*{-0.5cm}
\centering
\includegraphics[clip,width=0.52
\textwidth,height=0.55\textheight,angle=0.] {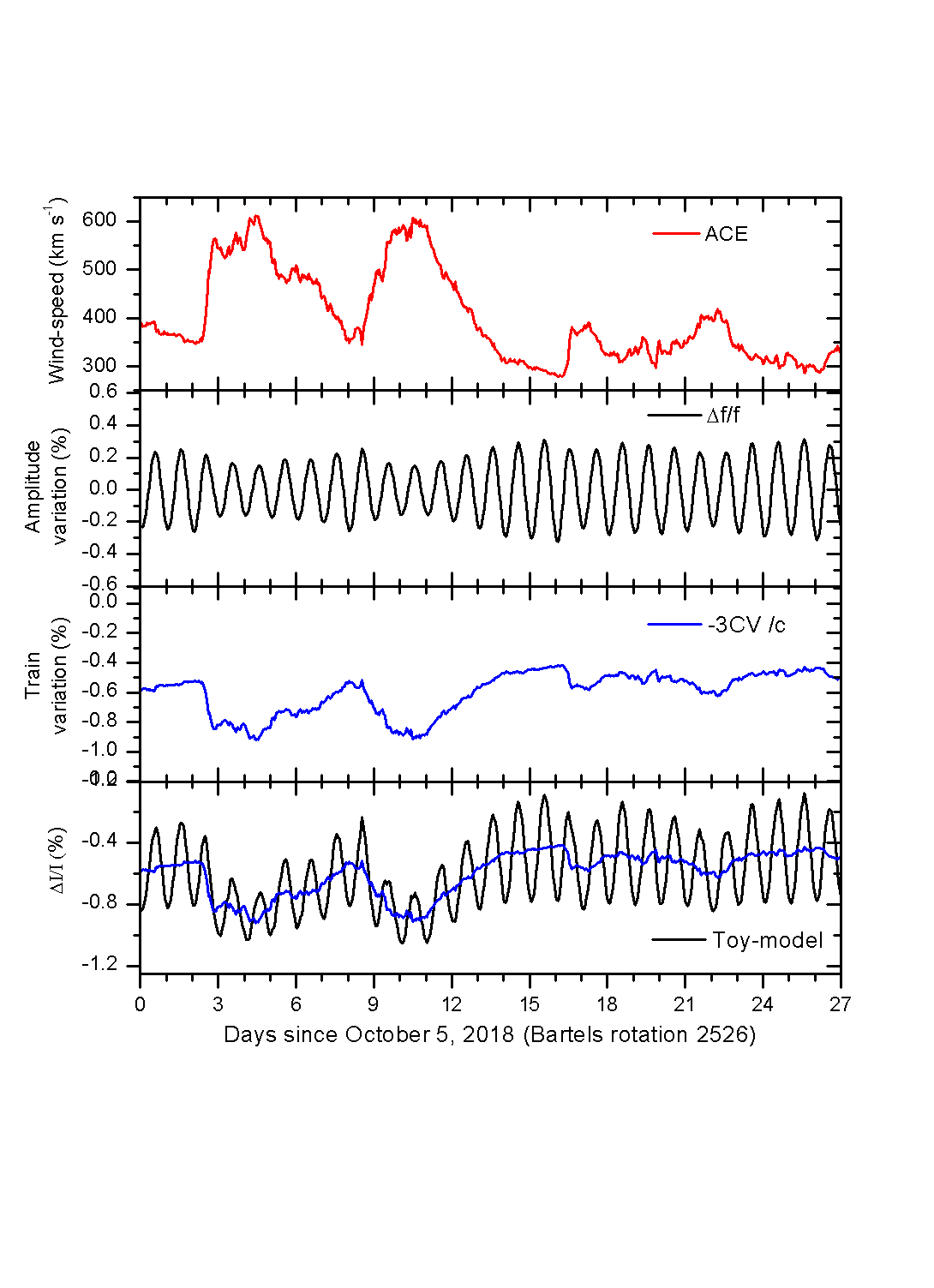}
\vspace*{-2.0cm}
\caption{Components of the Toy-model, from top to bottom.
First: ACE-SWEPAM solar wind hourly averages for the Bartels rotation 2526. Second: Solar wind modulation of the amplitude of the diurnal variation, according to Eq.~\ref{ani2}. Third: Solar wind modulation of the baseline of the diurnal variation train, first  term on the right of Eq.~\ref{toy}. Last panel: Particle hourly rates for the Bartels rotation 2526, according to Toy-model, Eq.~\ref{toy}.
}
\label{october_2019}
\end{figure} 


The average diurnal variation as observed by neutron
monitors over the period 1957-1965 \citep{mcc65,rau72,dugg67} is approximately constant over these years, with an amplitude of about  0.4\%. It is almost independent of the magnetic rigidity, but it has dependence with the latitude ($\lambda$), as $\cos \lambda$ \citep{rau72}. It means that at high latitudes (polar regions), the amplitude of the diurnal variation tends to disappear. At the PAO (latitude of -35.47 degrees), a diurnal variation remains even after atmospheric effects corrections with an average amplitude of 0.25\% 
\citep{auge11}, this value at the equator would be of  0.31\%. So the PAO result is around half of the predicted.

There are several hypotheses to reduce the amplitude of the diurnal variation predicted value concerning the observed, as the small cumulative errors in the calculated value, which together can reduce its value \citep{subr71}.
Moreover, according to \cite{joki69}, random fluctuations in the configuration of interplanetary magnetic field lines as a result of large-scale inhomogeneities in the solar wind, impose that drift processes cannot be ignored in GCR propagation.
The inclusion of drift dependence in the amplitude of the diurnal variation (Eq.~\ref{ani1}), is through a term in brackets \citep{levy76,erdo79}, and including the latitude dependence, as $\cos \lambda$,
the amplitude becomes
\begin{equation}
\frac{\Delta f}{f} \equiv 3C\frac{V}{c}\tan^2 \Psi \cos \lambda \left[ \frac{1-\alpha}{1+\alpha}\right],
\label{ani2}
\end{equation}
where $\alpha=K_{\perp}/K_{\parallel}$ is the ratio of perpendicular to parallel diffusion coefficients. In the limit of $\alpha << 1$, the drift effect is neglected and corresponds to a corotation of the cosmic ray flux with the Archimedes spiral field pattern. 
While, the other limit,$\alpha\lesssim 1$,
means $K_{\perp} \lesssim K_{\parallel}$, and the solar diurnal anisotropy vanishes.
So the galactic cosmic rays diffusion would almost isotropic.

The theoretical predicted average value of 0.6\% to the diurnal amplitude variation can be reduced to the average PAO observed value of 0.25\% if in Eq.~\ref{ani2} the PAO (Malargue) latitude, $\lambda$, is considered and a value of $\alpha=0.21$, to take into account the drift processes.

On the other hand, the second term on the right of Eq.~\ref{riker} take into account of the modulation
of the baseline of the diurnal variation train.
That term indicates that the modulation is proportional to solar-wind speed. For example,
when the Earth crosses a broad high-speed stream sector, with duration 
longer than the diurnal variation duration, the entire level of the cosmic ray train is modified and tends to follows the time profile of solar-wind speed.
However, the experimental data shows that the modification is in anti-correlation with the solar-wind
speed, meaning that this term is negative \citep{iucc79}. Thus, the cosmic ray intensity variation can be written as
\begin{equation}
\frac{\Delta I}{I}= -3C\frac{V}{c}+ \frac{\Delta f}{f} \sin (\frac{2 \pi}{T}t-\delta).
\label{toy}
\end{equation}
A harmonic function to generates the time profiles with a periodicity of $T=24$ h (Earth's rotation) was included.
$\delta$ is the phase, $\sim$18:00 h LT to $A<0$, or $\sim$15:00 h LT to $A>0$, where A is the magnetic polarity of the solar cycle. 

The above equation constitutes the semi-empirical Toy-model prediction. 
As an example, we show the components of the model. The top panel in 
Fig.~\ref{october_2019} shows the hourly averages of the solar-wind speed, according to ACE-SWEPAM solar-wind data from Bartels rotation 2526, with begin on October 5, 2018. We can see that in the period, there are two predominant HSSs, both with a maxima wind speed of about 600 km/s. 

The second panel shows the modulation of the amplitude of the diurnal variation of the GCR, according to Eq.~\ref{ani2}. 
Note that the amplitude is in anti-correlation with the solar wind-wind speed (see Fig.~\ref{corotating}).
The third panel represents the second term on the right of Eq.~\ref{riker}. It is the modulation of the diurnal variation train. The time behavior of this modulation is in inverse correlation with the time profile of the solar-wind speed, and means (as already commented) that this term is negative.
The fourth panel represents Eq.~\ref{toy}, which is the addition of the diurnal variation and the diurnal variation train. Eq.~\ref{toy} represents the essence of our Toy-model.

\section{Toy-Model validation}
\label{valitation}

\begin{figure}[th]
\vspace*{-0.0cm}
\hspace*{-0.5cm}
\centering
\includegraphics[clip,width=0.8
\textwidth,height=0.45\textheight,angle=0.] {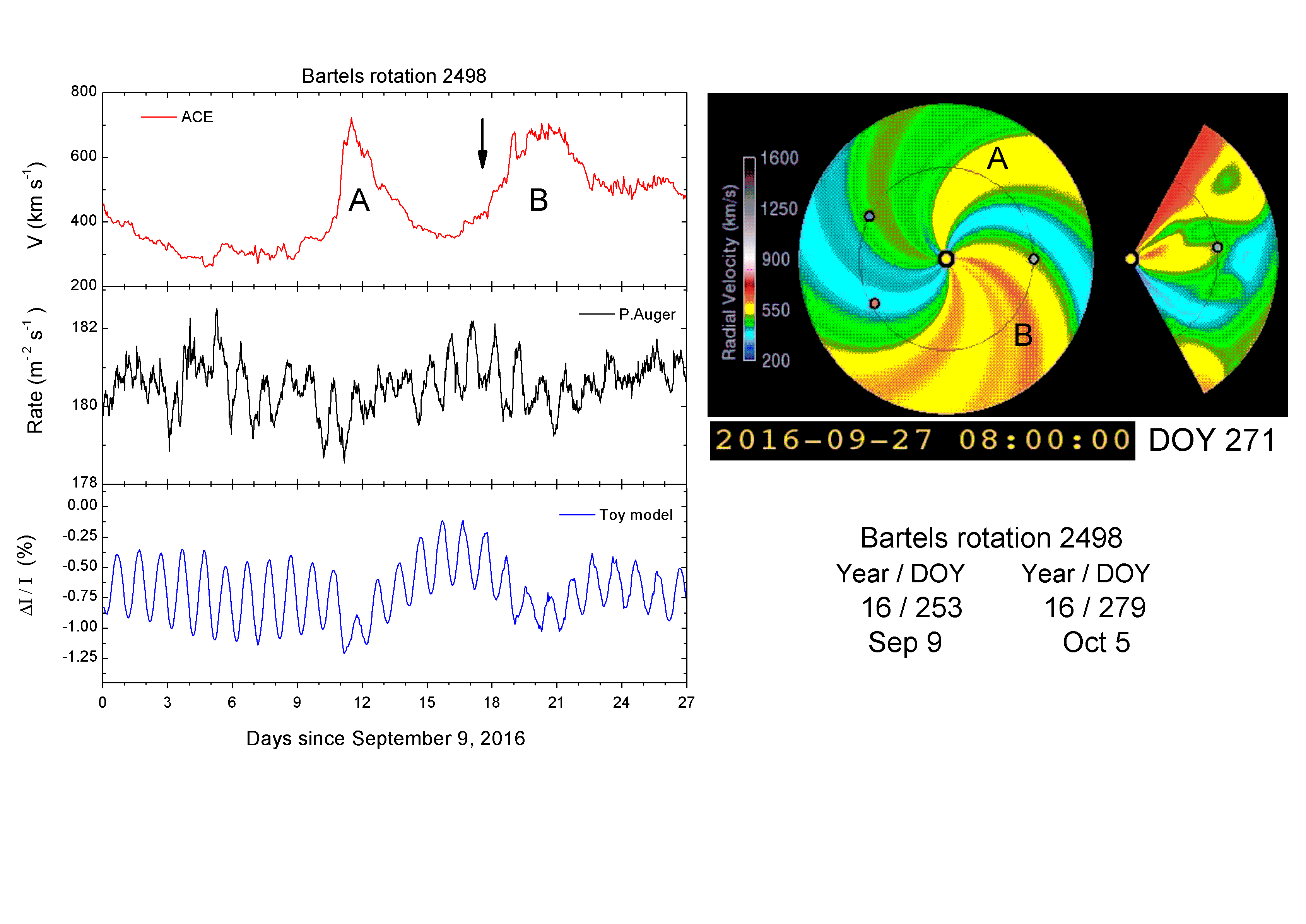}
\vspace*{-2.0cm}
\caption{Left panel: from top to bottom, hourly averages of the ACE-SWEPAM solar-wind speed at L1, 15-minutes averages from PAO escaler data, and the hourly averages from Toy-model, respectively, for the Bartels rotation 2498.  The arrow on the left represents the Earth's position on September 27, 2016.
Right panel: Real-time simulation of corotating and transient solar wind disturbances driven by the WSA-ENLIL prediction model, on September 27, 2016, at 08:00 UT. The two circles on the right represent the position of the two STEREO spacecraft, and the circle on the right represents the position of the Earth.
The labels A and B indicate two HSS structures. 
}
\label{september16auger}
\end{figure}

To show the accuracy of the Toy-Model, we made a qualitative and quantitative analysis, through a comparison with the data.
We focus on the Bartels rotation 2498, from September 9 to October 5, 2016. During this rotation, there were two predominant high-speed streams, both with speed reaching $\sim$ 600 km$s^{-1}$,  labeled as A and B in Fig.~\ref{september16auger} (right side), showing the WSA-Enlil prediction model, covering a part of this rotation. Also, on September 27, 2016, (DOY=271) at 08:00:00 UT, the Earth was at the edge of the HSS B after to crossed the HSS A.

Fig.~\ref{september16auger} (top panel) at the left side shows the hourly rates of the solar-wind speed, according to ACE-SWEPAM detector at Lagrange Point L1. The 15 minutes rates from PAO scaler mode data and the hourly rates according to Toy-Model are presented in the second and third panels, respectively. In all cases, the panels are for the full period of the Bartels rotation 2498 (2016). As was expected, the PAO time profiles, present fluctuations that are missing in the Toy-Model. However, the baseline modulation of the diurnal variation train in both, follow the solar wind variation.

The vertical arrow at left (top panel) indicates the Earth's position on September 27, 2016, after crosses the HSS A and about to cross the HSS B.

\begin{figure}[th]
\vspace*{-0.0cm}
\hspace*{-0.5cm}
\centering
\includegraphics[clip,width=0.37
\textwidth,height=0.5\textheight,angle=0.] {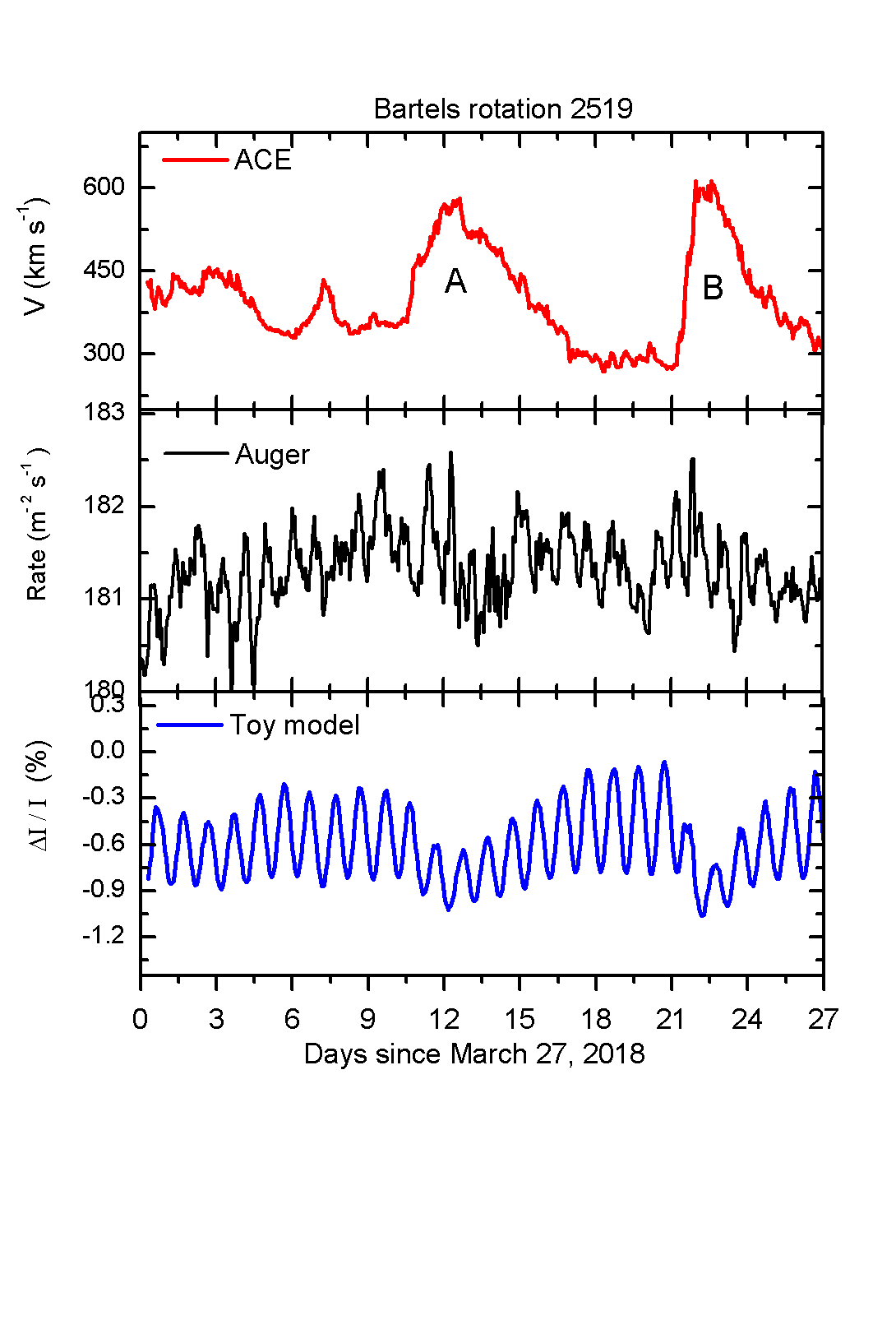}
\vspace*{-2.0cm}
\caption{Panels from top to bottom, hourly averages of the ACE-SWEPAM solar-wind speed at L1, of the PAO escaler data, and Toy-model, respectively, for the Bartels rotation 2519. 
}
\label{auger_march2018}
\end{figure}

Beyond a qualitative comparison, as shown above, a quantitative comparison can also be made looking, for example, if there is the same correlation between the variation in the count rate at the PAO and the toy model versus the solar wind variation. For this purpose, it is better also to put the data of PAO as hourly rates, and this is done for the Bartels rotation 2519. We believe that this will also improve the qualitative comparison, as shown in Fig.~\ref{auger_march2018}. We want to point out that the baseline of the diurnal variation train in the Toy-model depends on the solar wind speed observed at Lagrange Point L1. In general, a disturbance observed in L1 is seen on the ground-level, on average, three hours later. This behavior may explain the small time lag between PAO's data and the toy model.

As the time profiles of both solar wind speed and particle counting rates are time series. We use variations (in percentage) to compare the difference between two values that represent the same sort of variable, defined as $(A-\bar{A})/\bar{A}\times 100$, where $\bar{A}$ is the average value during the considered period.
Notice, the toy model output already provides this variation.

\begin{figure}[th]
\vspace*{-1.0cm}
\hspace*{-0.5cm}
\centering
\includegraphics[clip,width=0.6
\textwidth,height=0.43\textheight,angle=0.] {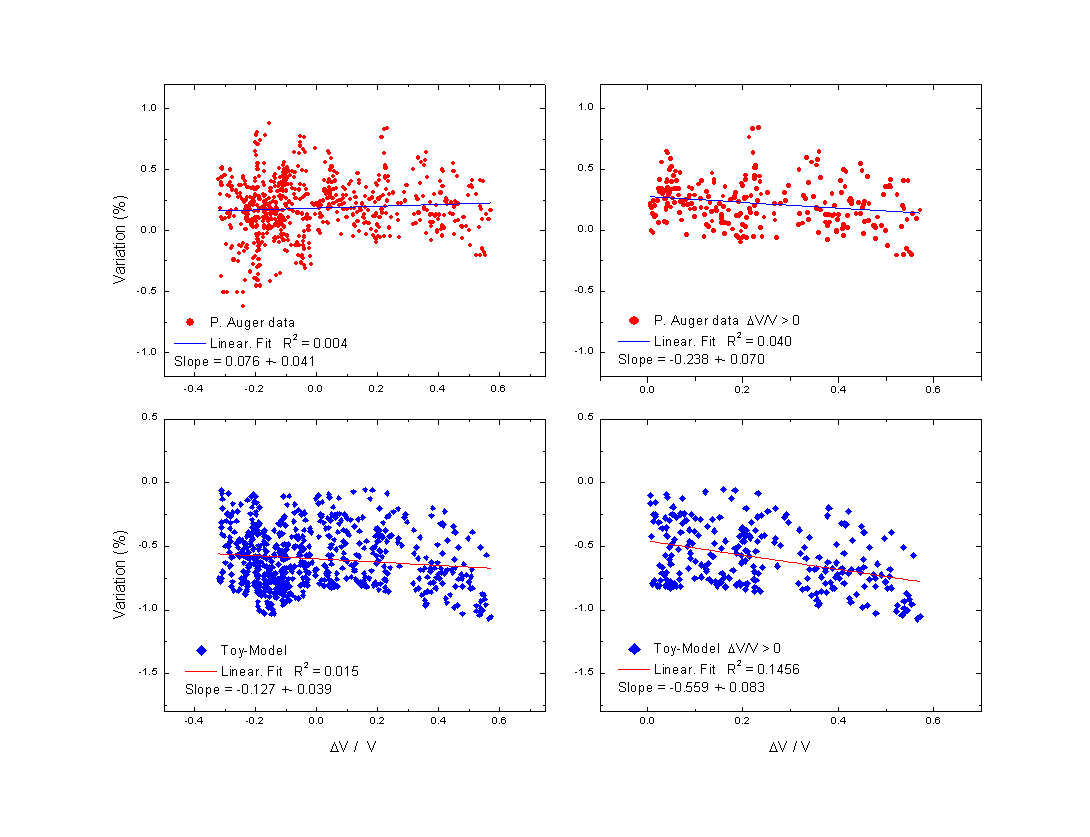}
\vspace*{-1.0cm}
\caption{Left panels: Histograms showing the anticorrelation between the hourly averages for the PAO scaler rates (top-panel) and the Toy-model (bottom panel), both with the solar wind variation,  for the  Bartels rotation 2498. Right panels: the same as in left panels, but after a cut and which includes only the points with $\Delta V/V>0$.
}
\label{auger_correlation}
\end{figure}

Fig.~\ref{auger_correlation} represents the hourly rates variation in the PAO's data (upper panels) and the variation of the hourly rate in the Toy-Model (lower panels), both with relation the solar wind speed variation, for the rotation of Bartels 2519, from March 27 to April 25, 2018.

Even considering that the Toy-Model does not take into account fluctuations, we can see that the relationship between the variation of the hourly rate in the toy model in relation with the hourly rates in the solar wind variation, is close to those observed in the 
 PAO's data, especially for the full Bartels rotation (left panels).
In both cases, the correlation parameter of a linear fit is close to zero shows a nonlinear trend. Even so, in both cases,  the values of the slopes are within the range of uncertainty. 

Also, in both cases, an anti-correlation between the variation in the counting rates and the solar wind speed variation is seen only after a cut, i.e., considering only the points where the solar-wind speed is higher than the average  value, or in other words, considering only
 positive values of the solar wind speed variation (panels on the right).
There is some discrepancy between the two slopes (PAO and Toy-model). Even so, they are still at the limit of the range of uncertainty.


\section{HSS from coronal mass ejection}

Solar flares originate from magnetic instabilities in groups of sunspots. An explosion generally triggers a coronal mass ejection (CME) into interplanetary space. If the explosion happens at low solar latitudes and not far from the central meridian, the  CME is geo-effective, depending on the ejection speed, after three days on average, the CME reaches the Earth. 
This phenomenon is dominant during the period of maximum solar activity (solar cycle of 11 years) and gradually decreases during the declining phase of the solar cycle.

CMEs are associated with strong solar wind disturbances.  
In most of the cases, the arrival of a CME trigger galactic cosmic rays depressions (Forbush decrease), observed at ground level detectors as a fall in the counting rate. Both the ejecta and shock associated with CME are responsible for the cosmic ray depressions \citep{cane93,cane94,cane03}.
Still, there is also contribution to the cosmic ray depressions associated with the solar wind disturbance (an increase of solar wind speed) at the front of the shock (HSS associated to a CME). If the CME is towards Earth, geomagnetic storms are expected.  Indeed, the most intense geomagnetic storms are due to the impact of a CME, reaching the level of up to G5 (extreme). However, the storms produced by HSSs from coronal holes are in most cases of level G1 (minor) and G2 (moderate) on the NOAA scale of geomagnetic storms.

\begin{figure*}[th]
\vspace*{-0.0cm}
\hspace*{-0.0cm}
\centering
\includegraphics[clip,width=0.9
\textwidth,height=0.5\textheight,angle=0.] {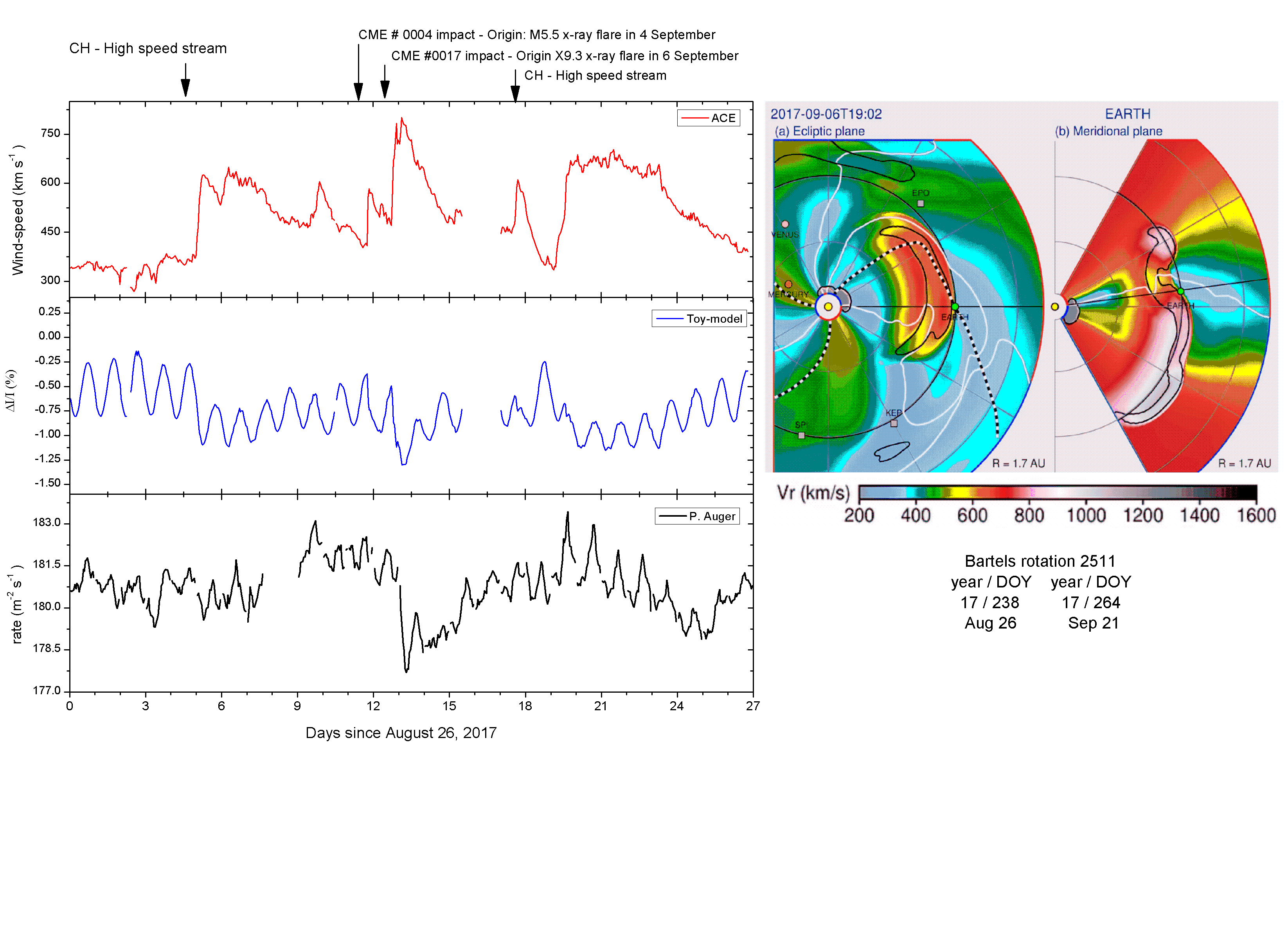}
\vspace*{-2.0cm}
\caption{Left panel: from top to bottom, hourly averages of the ACE-SWEPAM solar-wind speed at L1, of the Toy-model, and PAO escaler data, respectively, for the Bartels rotation 2511. Right panel:
Real-time simulation of corotating and transient solar wind disturbances driven by the WSA-ENLIL-DONKI-HELCATS model, on September 6, 2017, at 19:02 UT.
}
\label{september_2017}
\end{figure*} 

During the solar Bartels rotation 2511, from August 26, 2017, to September 21, 2017,
the magnetosphere was subject to the influence of several solar wind structures, such as the impact of up to two HSS from coronal holes, and two from CMEs. The vertical arrows indicate the sequence of these impacts on the top panel of Fig.~\ref{september_2017}.

Here we would like to highlight the effect of the solar wind structures from a CME on the cosmic rays at ground level.
 In the Bartels rotation 2511, two consecutive CMEs impacted the Earth, the first was during September 11, and the second one on September 12, they are indicated by the vertical arrows on the top of the Fig.~\ref{september_2017} (left panel). 
The second one had origin on the active solar area $\# AR12673$. The ejection triggered a  radiation storm (solar energetic particles, SEP) triggering a Ground Level Enhancement (GLE), it was the second on the solar cycle 24. Details on this event are available at \cite{augu19}.

The right panel in Fig.~\ref{september_2017} shows the propagation of these two solar wind structures (CMEs), according to the WSA-Enlil-DONKY model, available at \url{http://www.helioweather.net/archive/2017/09/}, when the first CME hits the Earth, on September 6, at 19:02 UT.
The left panel on the top of Fig.~\ref{september_2017} represents the hourly averages of solar wind speed from ACE-SWEPAM data.
 The middle and bottom panels on Fig.~\ref{september_2017} (left panels) show the Toy-Model prediction, 
 and the PAO data, respectively, in all cases, the data correspond to the Bartels rotation 2511.

 As the Toy-model takes into account only the solar wind disturbance in the cosmic ray depressions; a comparison between the PAO data, and the Toy-model, allows us to estimate the percentage of the intensity of the depression attributed to the HSS associated to the CME.
Following Fig.~\ref{september_2017} we can see that only about 39\% of the cosmic ray depression at PAO, in the transition of September 12 to 13 (impact of the second CME), is due to solar wind disturbance, so the remain 61\% is due to the impact of the bow-shock and plasma of the CME. These values may vary from event to event.

\section{Solar Cycle variation}

The solar wind speed variation is not uniform over the 11 years solar spots cycle. High-speed streams are present during the all solar cycle but tend to be more significant during the maximum solar cycle. The SOHO CELIAS data shows that the number of HSSs per Bartels rotation reaches up to six during the maximum solar cycle 23 period. But this number decreases up to three during the solar minimum, as shown in the lower panel of Fig.~\ref{soho}.

During the solar cycle maximum, in addition to the coronal holes, solar flares have a contribution to the generation of HSSs. While, during the declining phase,
the number the solar flares decreases and goes practically to be zero at the minimum of the solar cycle. So during this epoch, the HSSs from coronal holes turn dominant. Also, the coronal hole that originates de HSS is more stable during this epoch and survive during several Bartels rotations, so the HSSs are recurrent events with a periodicity of $\sim$27 days.

As above shown, the intensity of the cosmic rays is in anti-correlation with solar wind speed. Or in other words, the speed of the solar wind modulates the intensity of the cosmic rays, and the modulation has at least two components: the modulation of the amplitude of the diurnal variation and the modulation of the baseline of the diurnal variation train. The latter also is dominant in a long-term modulation analysis.

\begin{figure}[th]
\vspace*{-0.0cm}
\hspace*{-0.0cm}
\centering
\includegraphics[clip,width=0.5
\textwidth,height=0.3\textheight,angle=0.] {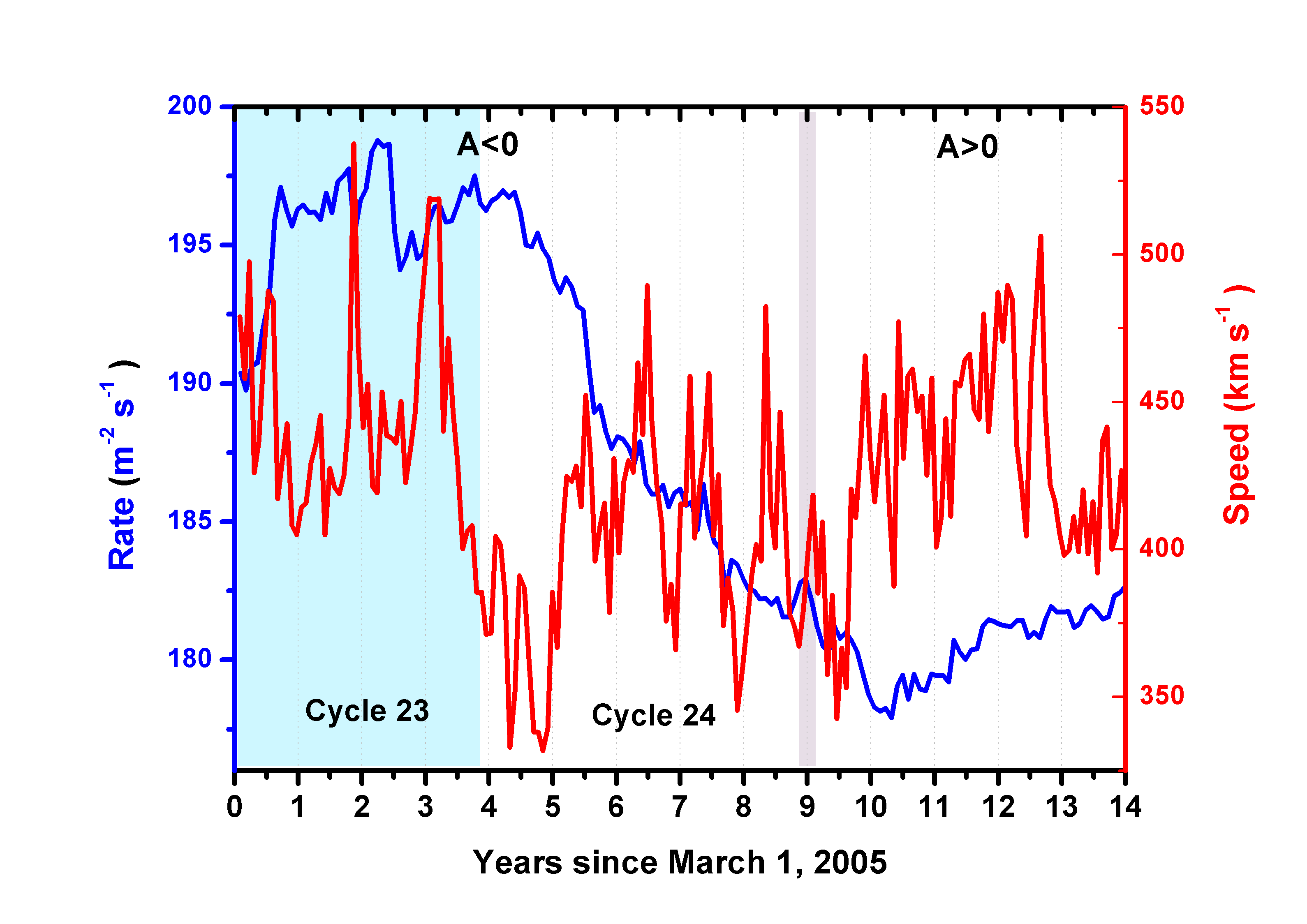}
\vspace*{-0.0cm}
\caption{The Bartels Rotation (27-day) averages, to a period of almost 15 years. 
The blue left scale and blue curve represent the PAO data. The red right scale and red curve represent the ACE-SWEPAM solar wind data.
}
\label{wind_total}
\end{figure} 

In fact, in almost 13 years of continuous PAO data, starting in March 2005, we can already see an anti-correlation
between the cosmic ray count rate in the surface array detectors and the solar wind speed recorded by the ACE probe at L1.
Fig.~\ref{wind_total} summarize the situation, and shows
the averages rates by Bartels rotation, of the PAO data, and the ACE solar wind speed data. Following the figure, we can see that the data cover the period of transition between solar cycles 23 to 24 (2008-2009). The light blue region corresponds to Cycle 23. As expected in this transition, the solar wind speed passes by a minimum, whereas the particle counting rate passes by a maximum.

Also, the data cover the period of maximum solar activity of the solar cycle 24. Besides weak, cycle 24 had two maxima, on $\sim$2012 and $\sim$2014. Even so, the anti-correlation is evident, the minimum of the intensity of the cosmic rays is close to the second increase of the solar wind speed and coincides with the change in the Sun magnetic polarity, from negative ($A<0$) to positive ($A>0$).

\section{Conclusions}
\label{conclusion}

Solar activity is known to influence galactic cosmic rays. This influence increases as the energy of cosmic rays decreases. At 1 AU, there are cosmic ray gradients (E$<$50TeV), transverse and perpendicular to the ecliptic plane, responsible for various types of anisotropies, such as diurnal, semi-diurnal and others on rotating Earth. Also,  there is an influence of solar corotating structures, such as the magnetic field forming a Parker spiral in the Ecliptic plane. As well as the solar wind, from coronal holes forming corotating HSS structures, exerting a modulation in cosmic radiation. Solar wind high-speed streams modulate the amplitude of the diurnal variation and its baseline variation train, observed on the particle counting rate at ground level. In short, the solar wind speed is in anti-correlation with the intensity of the cosmic rays.

In this paper, we have presented an analysis of the short-term modulation (one Bartels rotation) of the galactic cosmic rays 
(E$<$50 TeV) by the solar wind. The study is using the scaler mode data of the surface array detectors of the PAO, and the solar wind data from ACE-SWEPAM detector at L1.
We show that the modulation is consistent with at least two components. The first is the amplitude modulation of the diurnal variation, and the second one is the modulation on the baseline of the train of diurnal variation. Both modulations are in anti-correlation with solar-wind speed. 

We also show that these modulations can be described by the cosmic ray diffusion theory, more appropriately, the radial-gradient of the theory. But requires some extra assumptions,  as the dependence on the latitude of the diurnal variation, 
and the inclusion of drift processes in the galactic cosmic ray propagation.

Under these assumptions, we build a semi-empirical Toy-model. The main limitation of this model is that it does not incorporate fluctuations due to particle propagation into the atmosphere. We believe this step can be introduced later into a Monte Carlo simulation, and obtain more robust results.

The Toy-model reproduces the observations appropriately. They are showing through a qualitative and quantitative validation of the model through comparison with the PAO data.

\section{Acknowledgments}

 We express our gratitude to the Pierre Auger Observatory and the Space Weather Prediction Center (NOAA) for valuable information and data used in this study. 
 This work is supported by the National Council for Research (CNPq) of Brazil, under Grant 306605/2009-0, 402846/2016-8 and 301368/2019-8.

\end{document}